# Photometric observations of stars in the field of VV Cam and VW Cam

BOTTERO FRANCESCA[1], CARROZZA ALESSIO[1], FINI EDOARDO[1],
FINI VITTORIO[1], RAGO MARTA[1], SAVIO JACOPO[1] AND TRUSCELLI MARIA GRAZIA[1]
BENNA CARLO[2], GARDIOL DANIELE[2] AND PETTITI GIUSEPPE[2]

1) IIS Curie Vittorini, Corso Allamano 130, 10095, Grugliasco (TO), Italy, TOIS03400P@istruzione.it

2) INAF-Osservatorio Astrofisico di Torino, via Osservatorio 20, I-10025 Pino Torinese (TO), Italy, pettitg@alice.it

**Abstract:** the results of recent photometric observations of suspected variables stars VV Cam and VW Cam and of fainter stars in their field of view are reported and compared to data available from scientific literature and public astronomical databases.

## 1 Introduction

This report provides the results of our $BVR_cI_c$ photometric observations of suspected variable stars VV Cam and VW Cam and fainter stars in their field of view, carried on at Loiano site of the Astronomical Observatory of Bologna (OABo). The findings are compared with the outcome of previous jobs and data available from the ASAS-SN and NSVS databases.

VV Cam and VW Cam are part of a group of variable stars currently under investigation at the INAF-Astrophysical Observatory of Torino (OATo) and are classified as constant in the General Catalogue of Variable Stars (GCVS 5.1) and in the AAVSO databases.

After the initial observations and analysis (Nekrasova 1938 and Kukarkin 1948), a very few data are available on these two stars and their variability is not confirmed yet. The later observations indicate VV Cam as a possible variable and VW Cam as a constant star (Schmidt, 1992). A spectral type G5 III: was estimated for VV Cam (Bond 1978).

## 2 Observations

The instrumentation characteristics of the Cassini telescope used at Loiano site of the Bologna Astronomical Observatory (OABo) are reported in Table 1.

| Loiano site of the Bologna Astronomical Observatory (OABo) | | | | | | |
|---|---|---|---|---|---|---|
| **Telescope (Cassini)** | | | **Detector (BFOSC)** | | | |
| **Useful Diameter** | **Focal Length** | **Optical conf.** | **Camera** | **Array (pixels)** | **Johnson-Kron-Cousins** | **FoV** |
| 150.0 cm | 1200 cm | Ritchey-Chrétien | EEV D129915 | 1300x1340 | B, V, $R_c$, $I_c$ | 13'x12.6' |

**Table 1- Loiano site instrumentation characteristics**

The number of observations and length of exposures in each filter is shown in Table 2.

| | | Number of observations - Length of exposure (s) | | | |
|---|---|---|---|---|---|
| **Date** | **Time span** | **B** | **V** | **$R_c$** | **$I_c$** |
| 8 Dec, 2016 | 1h 37min | 30 - 15s | 30 - 10s | 30 - 5s | 30 - 5s |

**Table 2- Observation log**

The equatorial coordinates of the two suspected variables, at epoch J2000, are the following:
- VV Cam     R.A.    04h 59m 17.9712s        Decl.   +66° 19' 15.219"
- VW Cam     R.A.    05h 00m 44.9675s        Decl.   +66° 26' 56.961"

A finding chart with the identification of the two suspected variables, the comparison (C) and the check (K) stars is provided in Figure 1.

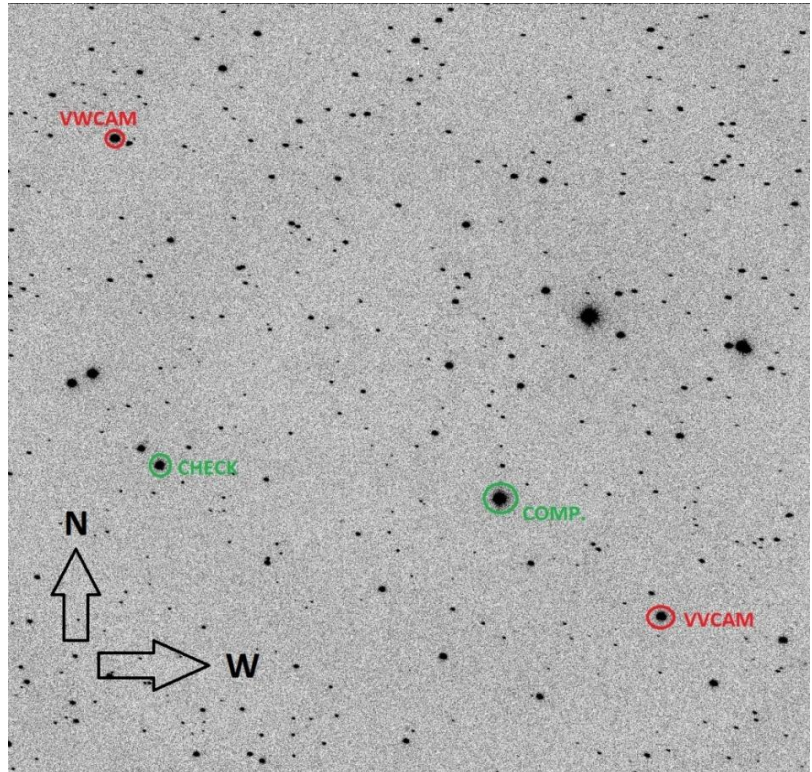

**Figure 1- VV Cam, VW Cam, Comparison and Check stars.**

The coordinates and the magnitudes of the comparison and check stars used for the photometry of VV Cam and VW Cam are shown in Table 3. For the comparison star the magnitudes are derived from the AAVSO APASS DR9 database: the magnitude in the filters $R_c$ and $I_c$ is calculated as mean value of the results of several transformations between g'r'i' and $R_c I_c$ filters applying the relation defined in Jester et al. (2005), Jordi et al. (2005) and Lupton (2005).

| Star | ID | RA [h m s] J2000.0 Gaia DR2 | Dec [° ´ ´´] J2000.0 Gaia DR2 | B (±3σ) | V (±3σ) | $R_c$ (±3σ) | $I_c$ (±3σ) |
|---|---|---|---|---|---|---|---|
| Comparison (C) | TYC 4091-1024-1 | 04 59 43.8224 | +66 21 09.857 | 12.188±0.077 | 11.154±0.070 | 10.669±0.098 | 10.223±0.068 |
| Check (K) | GSC 04091-01022 | 05 00 37.7845 | +66 21 42.348 | 13.825±0.080 | 13.010±0.070 | 12.610±0.100 | 12.174±0.070 |

**Table 3 - Comparison and check star data for VV Cam and VW Cam photometry**

We used the same star (C) to determine the magnitude of the check star shown in Table 3 and of additional four bright comparison stars (from C1 to C4) shown in Table 4 and Figure 2.

We could therefore build a set of six comparison stars (C, K, C1, C2, C3 and C4) that we used to perform photometry of fainter stars in the field of VV Cam and VW Cam, with the aim to search for short term variability. The use of six comparison stars allowed to determine the magnitudes of the fainter stars with a better accuracy.

| Star | ID | RA [h m s] J2000.0 Gaia DR2 | Dec [° ´ ´´] J2000.0 Gaia DR2 | B (±3σ) | V (±3σ) | $R_c$ (±3σ) | $I_c$ (±3σ) |
|---|---|---|---|---|---|---|---|
| C1 | GSC 04091-01031 | 05 00 40.7712 | +66 21 57.818 | 14.61±0.08 | 13.85±0.07 | 13.48± 0.10 | 13.08±0.07 |
| C2 | GSC 04091-01076 | 05 00 51.8324 | +66 23 00.976 | 13.77±0.08 | 12.82±0.07 | 12.35±0.10 | 11.86±0.07 |
| C3 | TYC 4091-1087-1 | 05 00 48.5569 | +66 23 09.996 | 12.91±0.08 | 12.20±0.07 | 11.86±0.10 | 11.48±0.07 |
| C4 | TYC 4091-1171-1 | 04 59 29.3680 | +66 24 06.812 | 11.32±0.08 | 10.37±0.07 | 9.90±0.10 | 9.49±0.07 |

**Table 4 - Additional comparison stars**

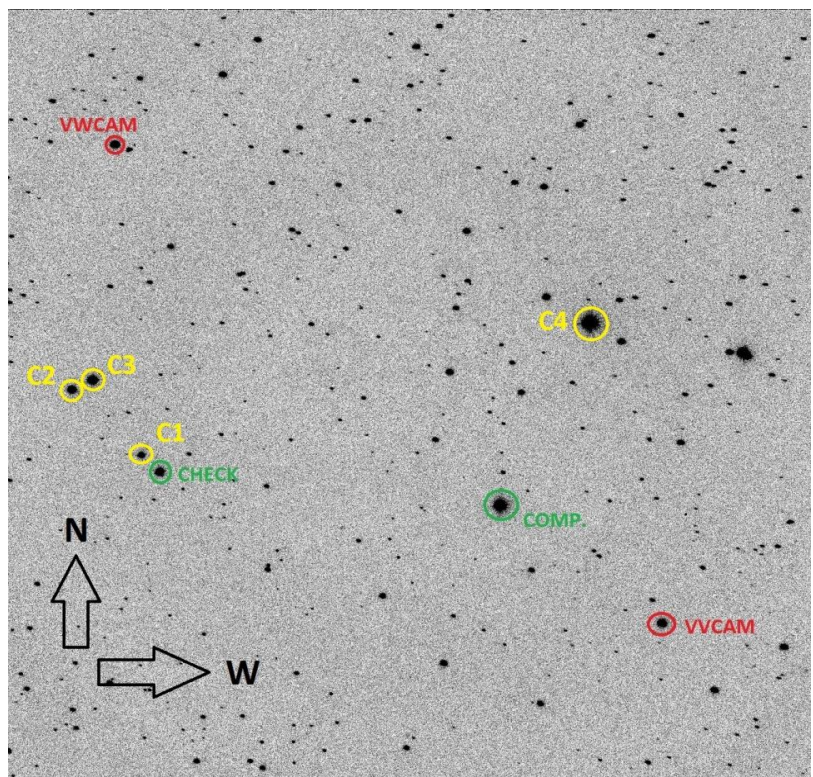

**Figure 2 - Additional comparison stars in the field of VV Cam and VW Cam**

The photometry of the suspected variables and of the comparison stars was performed using an IDL software developed at OATo whilst the photometry of the fainter stars has been performed using AstroArt 5.0.

## 3   Data analysis

For all images, Dark and Flat Field corrections were applied, and aperture differential photometry was performed.

We used the basic equation to obtain standard magnitudes from instrumental magnitudes in the form (for filter V) shown in Eq. 1:

$$V_{var} = \Delta v + T_v \cdot \Delta(B-V) + V_{comp} \qquad (1)$$

where:
- $\Delta v$ is the instrumental magnitude of the variable minus the instrumental magnitude of the comparison star;
- $V_{comp}$ is the V–magnitude of the comparison stars defined in Tables 3 and 4;
- $T_v$ is the transformation coefficient defined in Table 5;
- $\Delta(B-V)$ is the difference between the standard color of the variable and the standard color of the comparison stars, computed using the formula:

$$\Delta(B-V) = T_{bv} \cdot \Delta(b-v)$$

being $\Delta(b-v)$ the difference between the instrumental color of the variable and the instrumental color of the comparison stars.

Similar equations were applied for the filters B, $R_c$, $I_c$.

| $T_b$ |   | 0.188 ± 0.013 | $T_r$ | -0.060 ± 0.014 |
|---|---|---|---|---|
|   | for B, V | -0.034 ± 0.007 | $T_i$ | 0.033 ± 0.007 |
| $T_v$ | for $R_c$ | -0.067 ± 0.015 | $T_{bv}$ | 1.284 ± 0.008 |
|   | for $I_c$ | -0.034 ± 0.008 | $T_{vr}$ | 1.000 ± 0.011 |
|   |   |   | $T_{vi}$ | 0.938 ± 0.008 |

**Table 5 - Transformation coefficients**

Observed data falling outside the range of ±3σ from the mean values have been deleted.

The observed C-K delta magnitudes between the comparison and the check stars do not show significant trends or discontinuities, and the computed errors associated to the instrumental magnitudes in each filter are compatible with the observed C-K dispersion. However, the error associated to the transformed standard magnitudes is dominated by the uncertainty of the comparison star.

## 4    Results

The mean value of the standard magnitudes, color indexes and HJD calculated from our photometric observations of VV Cam and VW Cam are shown in Table 6. All $BVR_cI_c$ standard measures are available in Appendixes 1 and 2.

| Star | RA [h m s] J2000.0 Gaia DR2 | Dec [° ´ ´´] J2000.0 Gaia DR2 | B (±3σ) | V (±3σ) | $R_c$ (±3σ) | $I_c$ (±3σ) | B-V (±3σ) | $V-R_c$ (±3σ) | $V-I_c$ (±3σ) |
|---|---|---|---|---|---|---|---|---|---|
| VV Cam | 04 59 17.9712 | +66 19 15.219 | 13.66 ±0.09 | 12.70 ±0.08 | 12.24 ±0.10 | 11.78 ±0.07 | 0.96 ±0.11 | 0.46 ±0.12 | 0.92 ±0.10 |
| VW Cam | 05 00 44.9675 | +66 26 56.961 | 14.09 ±0.09 | 12.90 ±0.08 | 12.32 ±0.10 | 11.73 ±0.08 | 1.19 ±0.11 | 0.58 ±0.13 | 1.17 ±0.11 |
| | | Mean HJD (2457730 +) | 0.69755 | 0.69817 | 0.69869 | 0.69923 | 0.69786 | 0.69843 | 0.69870 |

**Table 6 - VV Cam and VW Cam BVRI data**

The known J, H, K magnitudes of the two stars (Cutri et al. 2003) and the color indexes V-J, V-H and V-K derived using our mean V magnitude are reported in Table 7.

| Star | RA [h m s] J2000.0 Gaia DR2 | Dec [° ´ ´´] J2000.0 Gaia DR2 | J | H | K | V-J (±3σ) | V-H (±3σ) | V-K (±3σ) |
|---|---|---|---|---|---|---|---|---|
| VV Cam | 04 59 17.9712 | +66 19 15.219 | 11.170 ±0.023 | 10.821 ±0.028 | 10.714 ±0.021 | 1.53 ±0.08 | 1.88 ±0.08 | 1.99 ±0.08 |
| VW Cam | 05 00 44.9675 | +66 26 56.961 | 10.941 ±0.027 | 10.407 ±0.031 | 10.253 ±0.019 | 1.96 ±0.08 | 2.49 ±0.09 | 2.65 ±0.08 |

**Table 7 - VV Cam and VW Cam JHK data**

## 4.1 Short Term Variability of VV Cam e VW Cam

In the time span covered by our measurements, no short-term variability was observed for the two stars, as shown by Figures 3 and 4, that report the standard magnitude in the V filter against time (HJD-2457730).

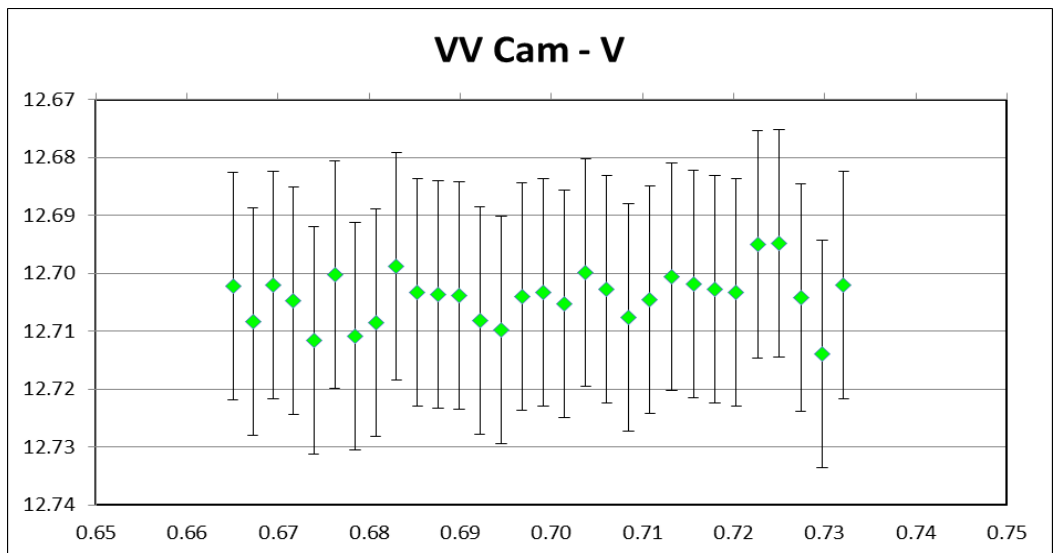

**Figure 3 - Standard Magnitude V vs HJD for VV Cam**

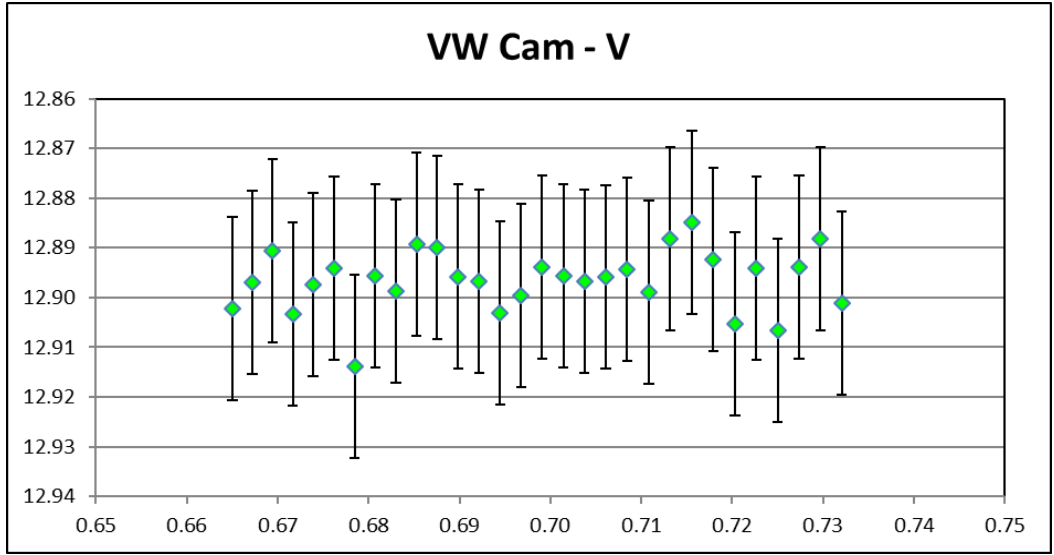

**Figure 4 – Standard Magnitude V vs HJD for VW Cam**

## 4.2 Long Term Variability of VV Cam and VW Cam

To investigate the long-term behavior of the two stars, we analyzed our results together with the data available from Schmidt (1992), and from the ASAS-SN and NSVS databases. Table 8 summarizes the mean magnitudes from the available observations.

| VV Cam | | | | | |
|---|---|---|---|---|---|
| | **Mean V mag (±1σ)** | **Mean R mag (±1σ)** | **B-V (±1σ)** | **V-R$_c$ (±1σ)** | **V-I$_c$ (±1σ)** |
| **OATo (2016)** | 12.70 ± 0.03 | 12.24 ± 0.03 | 0.96 ± 0.04 | 0.46 ± 0.04 | 0.92 ± 0.03 |
| **Schmidt (1992)** | 12.77 ± 0.01 | 12.29 ± 0.01 | | 0.48 ± 0.01 | |
| **ASAS-SN** | 12.730 ± 0.014 | | | | |
| VW Cam | | | | | |
| **OATo (2016)** | 12.90 ± 0.03 | 12.32 ± 0.03 | 1.19 ± 0.04 | 0.58 ± 0.04 | 1.17 ± 0.04 |
| **Schmidt (1992)** | 12.93 ± 0.01 | 12.35 ± 0.01 | | 0.58 ± 0.01 | |
| **ASAS-SN** | 12.930 ± 0.016 | | | | |

**Table 8 - Summary of available measurements for VV Cam and VW Cam**

Figures from 5 to 8 show the ASAS-SN measurements of the two stars in the period from January 2012 to April 2018. NSVS measurements are reported in Figures 9 and 10.

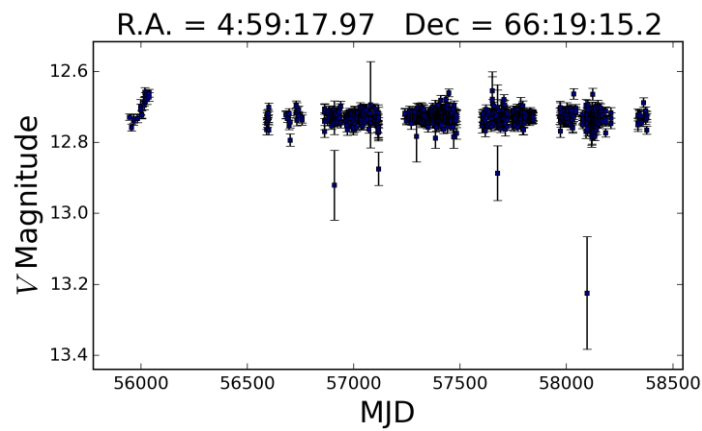

**Figure 5 - VV Cam ASAS-SN measurements**

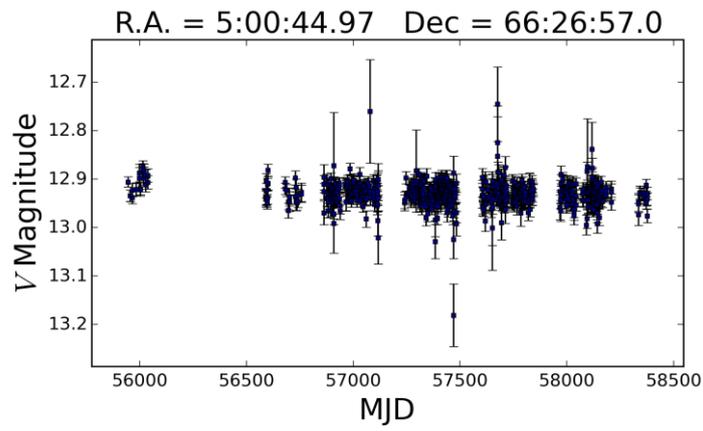

**Figure 6 - VW Cam ASAS-SN measurements**

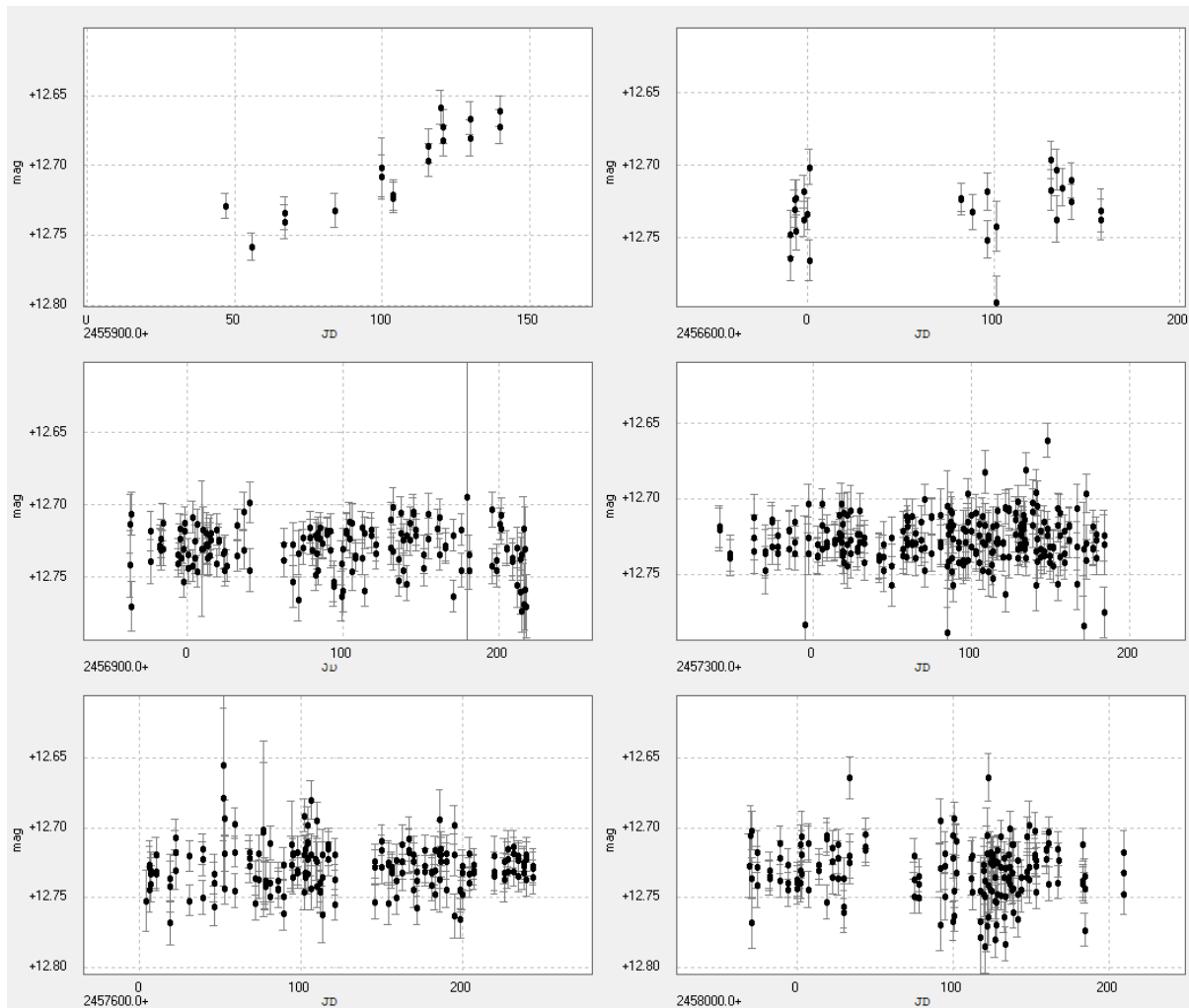

**Figure 7 – VV Cam ASAS-SN measurements expansion**

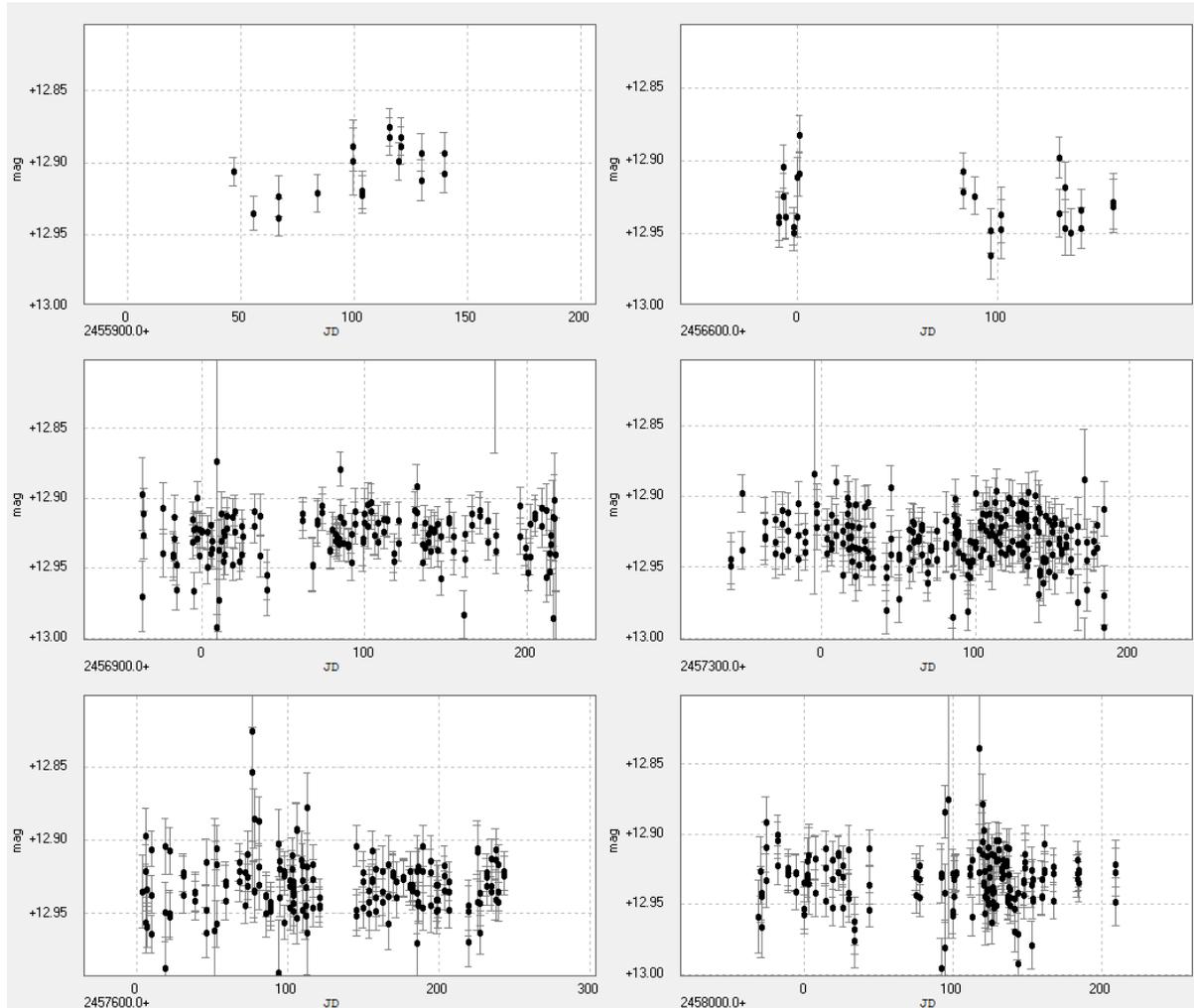

**Figure 8 – VW Cam ASAS-SN measurements expansion**

From the analysis of the ASAS-SN photometric data, considering that the variation of luminosity appears significantly greater than the 3σ error of the measurements, that is $0.04^m$, we have concluded that the variability of VV Cam and VW Cam cannot be excluded, and further observations are needed.

Moreover, even if the two stars generally do not show a preference for the maximum or the minimum level of luminosity, for VV Cam, a slow but regular increase of $0.10^m$ was observed in the period from HJD 2455955 to 2456040. Unfortunately, the lack of data in the subsequent 1.5-year period, does not allow to investigate the evolution of this event. Similar events did not occur in the time-span covered by the NSVS observations (Wozniak, 2004).

From all photometric data available, the amplitude for VV Cam is $0.13^m$ in the filter V, with a typical observed magnitude in the range 12.66 ÷ 12.79 and a mean value of 12.730 ± 0.014 (1σ). For VW Cam the amplitude is $0.11^m$ in the filter V, in a range between 12.88 ÷ 12.99 and a mean value of 12.930 ± 0.016 (1σ).

Measurements outside the above ranges are observed but it is noted that are all associated to observations with greater errors and worse magnitude limits and are considered not significative.

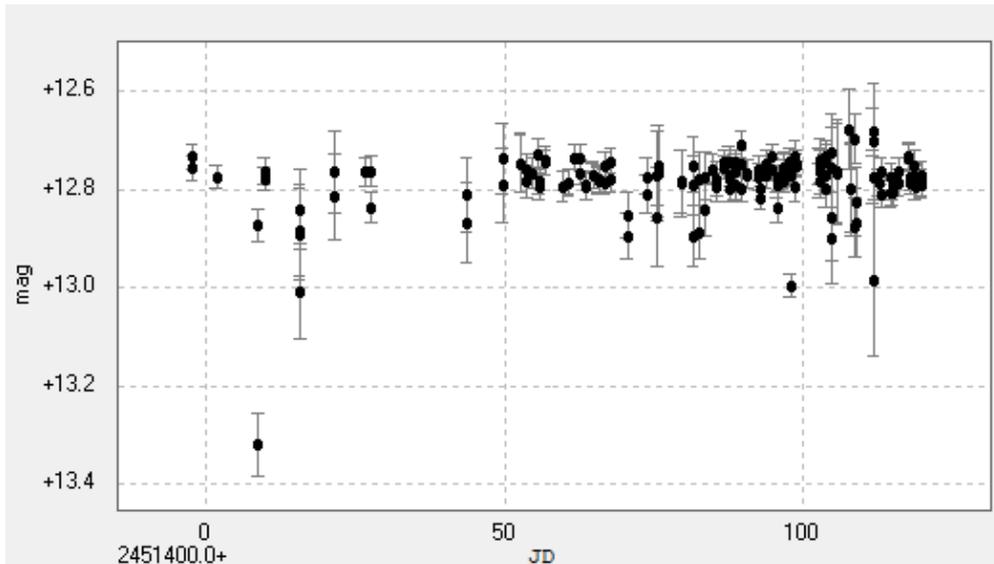

**Figure 9 – VV Cam ROTSE mag vs JD**

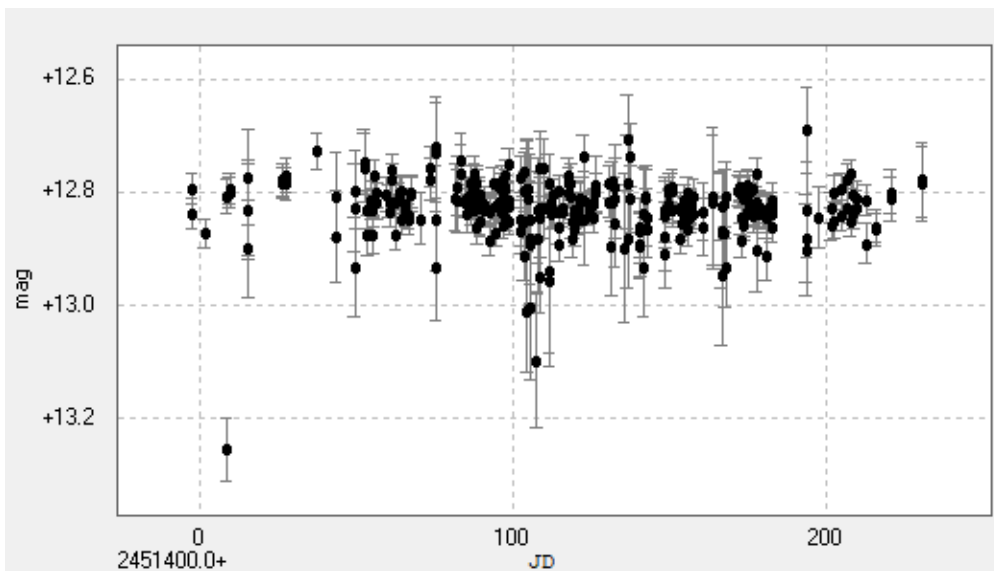

**Figure 10 – VW Cam ROTSE mag vs JD**

For both variables, ASAS-SN and NSVS measurements were analyzed using the version 2.51 of the light curve and period analysis software PERANSO (Paunzen, E., Vanmunster, T. 2016), but no significant periods were found.

## 4.3 Interstellar extinction and intrinsic magnitudes

To calculate the intrinsic magnitudes, color indexes and standard absolute magnitude $M_V$ in filter V of the two variables reported in Table 10, we used the mean value of our measurements, summarized in Table 6 and the JHK magnitudes in Table 7, corrected using the interstellar extinction coefficients $A_\lambda$ for the filters available from the NASA/IPAC Extragalactic Database, from E. Schlafly et al. (2011ApJ...737..103S) and the parallax *p* available from Gaia DR2 release (Luri 2018), reported in Table 9.

|  | Interstellar extinction $A_\lambda$ (mag/kpc) | | | | | | | parallax (mas) | distance (pc) |
|---|---|---|---|---|---|---|---|---|---|
|  | *B* | *V* | $R_c$ | $I_c$ | *J* | *H* | *K* | *p* | *d* |
| **VV Cam** | 0.904 | 0.683 | 0.541 | 0.375 | 0.177 | 0.112 | 0.075 | 1.9951 ± 0.0261 | 501 ± 7 |
| **VW Cam** | 0.765 | 0.578 | 0.458 | 0.317 | 0.150 | 0.095 | 0.064 | 0.4802 ± 0.0251 | 2082 ± 115 |

**Table 9: extinction coefficients for VV Cam and VW Cam**

|  | Intrinsic magnitudes | | | | Intrinsic color indexes | | |
|---|---|---|---|---|---|---|---|
|  | *B* | *V* | $R_c$ | $I_c$ | *B-V* | $V-R_c$ | $V-I_c$ |
| **VV Cam** | 13.21 ± 0.09 | 12.36 ± 0.08 | 11.97 ± 0.10 | 11.59 ± 0.07 | 0.85 ± 0.12 | 0.39 ± 0.13 | 0.77 ± 0.11 |
| **VW Cam** | 12.50 ± 0.09 | 11.70 ± 0.08 | 11.37 ± 010 | 11.07 ± 0.08 | 0.80 ± 0.12 | 0.33 ± 0.13 | 0.63 ± 0.11 |
|  | *J* | *H* | *K* | *V-J* | *V-H* | *V-K* | $M_V$ |
| **VV Cam** | 11.081 ± 0.023 | 10.765 ± 0.028 | 10.676 ± 0.021 | 1.28 ± 0.08 | 1.59 ± 0.08 | 1.68 ± 0.08 | 3.86 ± 0.08 |
| **VW Cam** | 10.866 ± 0.027 | 10.359 ± 0.031 | 10.221 ± 0.019 | 0.83 ± 0.08 | 1.34 ± 0.09 | 1.48 ± 0.08 | 0.10 ± 0.08 |

**Table 10: intrinsic magnitudes, color indexes and absolute magnitude for VV Cam and VW Cam**

## 4.4 Spectral type determination from photometric characteristics

We estimated the spectral types and the luminosity class of the two variables comparing the intrinsic color indexes and absolute magnitude in the filter *V*, defined in Table 11 with the characteristics derived by Pickels (1998).
As shown in Table 11, the photometric characteristics of the stars are consistent:
- with a G5 IV- G6 IV subgiant star for VV Cam;
- with a F5 III - F8 III giant star for VW Cam.

The color indexes in italics are interpolated from the original data available by Pickels (1998).

|  |  | $V-R_c$ | $V-I_c$ | *V-J* | *V-H* | *V-K* | $M_V$ |
|---|---|---|---|---|---|---|---|
|  | **VV Cam** | **0.39 ± 0.13** | **0.77 ± 0.11** | **1.28 ± 0.08** | **1.59 ± 0.08** | **1.68 ± 0.08** | **3.86 ± 0.08** |
| Pickels (1998) | G2 IV | 0.34 | 0.70 | 1.11 | 1.40 | 1.50 |  |
|  | G3 IV | *0.36* | *0.72* | *1.13* | *1.42* | *1.52* |  |
|  | G4 IV | *0.37* | *0.74* | *1.17* | *1.46* | *1.56* |  |
|  | G5 IV | **0.39** | **0.77** | *1.21* | *1.51* | *1.60* |  |
|  | G6 IV | *0.41* | *0.80* | ***1.26*** | ***1.57*** | ***1.66*** |  |
|  | G7 IV | *0.43* | *0.83* | *1.33* | *1.65* | *1.72* |  |
|  | G8 IV | 0.45 | 0.86 | 1.40 | 1.74 | 1.80 |  |
|  |  |  |  |  |  |  |  |
|  |  | $V-R_c$ | $V-I_c$ | *V-J* | *V-H* | *V-K* | $M_V$ |
|  | **VW Cam** | **0.33 ± 0.13** | **0.63 ± 0.11** | **0.83 ± 0.08** | **1.34 ± 0.09** | **1.48 ± 0.08** | **0.10 ± 0.08** |
| Pickels (1998) | F4 III | *0.24* | *0.48* | *0.75* | *0.99* | *0.97* |  |
|  | F5 III | 0.25 | 0.51 | **0.80** | 1.09 | 1.08 |  |
|  | F6 III | *0.27* | *0.54* | ***0.87*** | *1.20* | *1.20* |  |
|  | F7 III | *0.29* | *0.58* | *0.96* | ***1.33*** | *1.32* |  |
|  | F8 III | ***0.31*** | ***0.62*** | *1.06* | *1.46* | ***1.46*** |  |
|  | F9 III | ***0.34*** | *0.67* | *1.18* | *1.60* | *1.60* |  |
|  | G0 III | 0.37 | 0.73 | 1.31 | 1.74 | 1.76 |  |

**Table 11 – VV Cam and VW Cam photometric characteristics vs spectral types**

## 4.5 Photometric characteristics and spectral type of additional stars

The results of our photometric analysis of the comparison stars, defined in Tables 3 and 4, and of additional stars in the field of view of the two variables, are summarized in Tables 12. The JHK magnitudes (Cutri et al. 2003) and the color indexes are shown in Table 13 and 14 respectively. With the same approach used for VV Cam and VW Cam, we determined the intrinsic colors indexes and absolute magnitude of the stars. The interstellar extinction coefficients $A_\lambda$ and the parallax $p$ used for each and star are defined in Table 15; intrinsic magnitudes, color indexes and standard absolute magnitude $M_V$ in filter V are reported in Table 16. An estimation of the spectral types and class of luminosities of these stars is provided in Table 17, for the first time, based on their photometric characteristics. For none of the stars our photometry did show a significant sign of short term variability.

| Star | RA [h m s] J2000.0 Gaia DR2 | Dec [° ´ ´´] J2000.0 Gaia DR2 | B (±3σ) | V (±3σ) | $R_c$ (±3σ) | $I_c$ (±3σ) |
|---|---|---|---|---|---|---|
| **TYC 4091-1024-1** | 04 59 43.8224 | +66 21 09.857 | 12.188 ± 0.077 | 11.154 ± 0.070 | 10.669 ± 0.098 | 10.223 ± 0.068 |
| **GSC 04091-01022** | 05 00 37.7845 | +66 21 42.348 | 13.825 ± 0.080 | 13.010 ± 0.070 | 12.610 ± 0.100 | 12.174 ± 0.070 |
| **GSC 04091-01031** | 05 00 40.7712 | +66 21 57.818 | 14.61 ± 0.08 | 13.85 ± 0.07 | 13.48 ± 0.10 | 13.08 ± 0.07 |
| **GSC 04091-01076** | 05 00 51.8324 | +66 23 00.976 | 13.77 ± 0.08 | 12.82 ± 0.07 | 12.35 ± 0.10 | 11.86 ± 0.07 |
| **TYC 4091-1087-1** | 05 00 48.5569 | +66 23 09.996 | 12.91 ± 0.08 | 12.20 ± 0.07 | 11.86 ± 0.10 | 11.48 ± 0.07 |
| **TYC 4091-1171-1** | 04 59 29.3680 | +66 24 06.812 | 11.32 ± 0.08 | 10.37 ± 0.07 | 9.90 ± 0.10 | 9.49 ± 0.07 |
| **TYC 4091-1838-1** | 04 59 05.0149 | +66 23 35.980 | 12.37 ± 0.04 | 11.53 ± 0.06 | 11.13 ± 0.06 | 10.75 ± 0.03 |
| **GSC 04091-00302** | 04 59 36.1792 | +66 24 31.406 | 14.81 ± 0.13 | 13.72 ± 0.02 | 13.11 ± 0.03 | 12.56 ± 0.01 |
| **GSC 04091-01109** | 04 58 58.7218 | +66 18 52.154 | 14.65 ± 0.12 | 13.83 ± 0.03 | 13.40 ± 0.03 | 12.96 ± 0.02 |
| **GSC 04091-01163** | 04 59 24.3080 | +66 23 47.048 | 14.65 ± 0.23 | 13.85 ± 0.02 | 13.44 ± 0.04 | 13.02 ± 0.02 |

**Table 12 – OATo BVRI measurements of comparison and additional stars**

| Star | J | H | K |
|---|---|---|---|
| **TYC 4091-1024-1** | 9.573 ± 0.023 | 9.193 ± 0.026 | 9.077 ± 0.020 |
| **GSC 04091-01022** | 11.883 ± 0.042 | 11.627 ± 0.048 | 11.445 ± 0.033 |
| **GSC 04091-01031** | 12.638 ± 0.027 | 12.343 ± 0.032 | 12.305 ± 0.028 |
| **GSC 04091-01076** | 11.241 ± 0.027 | 10.893 ± 0.030 | 10.792 ± 0.021 |
| **TYC 4091-1087-1** | 11.028 ± 0.026 | 10.838 ± 0.031 | 10.742 ± 0.019 |
| **TYC 4091-1171-1** | 8.917 ± 0.020 | 8.510 ± 0.028 | 8.419 ± 0.023 |
| **TYC 4091-1838-1** | 10.381 ± 0.027 | 10.151 ± 0.031 | 10.081 ± 0.026 |
| **GSC 04091-00302** | 11.800 ± 0.024 | 11.316 ± 0.028 | 11.216 ± 0.022 |
| **GSC 04091-01109** | 12.406 ± 0.024 | 12.098 ± 0.029 | 12.007 ± 0.024 |
| **GSC 04091-01163** | 12.491 ± 0.025 | 12.192 ± 0.029 | 12.146 ± 0.024 |

**Table 13 – JHK data of comparison and additional stars**

| Star | B-V (±3σ) | V-R$_c$ (±3σ) | V-I$_c$ (±3σ) | V-J (±3σ) | V-H (±3σ) | V-K (±3σ) |
|---|---|---|---|---|---|---|
| **TYC 4091-1024-1** | 1.03 ± 0.10 | 0.48 ± 0.12 | 0.93 ± 0.10 | 1.58 ± 0.07 | 1.96 ± 0.07 | 2.08 ± 0.07 |
| **GSC 04091-01022** | 0.82 ± 0.11 | 0.40 ± 0.12 | 0.84 ± 0.10 | 1.13 ± 0.08 | 1.38 ± 0.08 | 1.57 ± 0.08 |
| **GSC 04091-01031** | 0.76 ± 0.11 | 0.37 ± 0.12 | 0.77 ± 0.10 | 1.21 ± 0.08 | 1.51 ± 0.08 | 1.55 ± 0.08 |
| **GSC 04091-01076** | 0.95 ± 0.11 | 0.47 ± 0.12 | 1.00 ± 0.10 | 1.68 ± 0.08 | 1.93 ± 0.08 | 2.03 ± 0.07 |
| **TYC 4091-1087-1** | 0.71 ± 0.11 | 0.34 ± 0.12 | 0.72 ± 0.10 | 1.17 ± 0.07 | 1.36 ± 0.08 | 1.46 ± 0.07 |
| **TYC 4091-1171-1** | 0.95 ± 0.11 | 0.47 ± 0.12 | 0.88 ± 0.10 | 1.45 ± 0.07 | 1.86 ± 0.08 | 1.95 ± 0.07 |
| **TYC 4091-1838-1** | 0.84 ± 0.07 | 0.40 ± 0.08 | 0.78 ± 0.07 | 1.15 ± 0.07 | 1.38 ± 0.07 | 1.45 ± 0.07 |
| **GSC 04091-00302** | 1.09 ± 0.13 | 0.61 ± 0.04 | 1.16 ± 0.02 | 1.92 ± 0.03 | 2.40 ± 0.03 | 2.50 ± 0.03 |
| **GSC 04091-01109** | 0.82 ± 0.12 | 0.43 ± 0.04 | 0.87 ± 0.04 | 1.42 ± 0.04 | 1.73 ± 0.04 | 1.82 ± 0.04 |
| **GSC 04091-01163** | 0.80 ± 0.23 | 0.41 ± 0.04 | 0.83 ± 0.03 | 1.36 ± 0.03 | 1.66 ± 0.04 | 1.70 ± 0.03 |

**Table 14 – Color indexes of comparison and additional stars**

| | Interstellar extinction A$_\lambda$ (mag/kpc) | | | | | | | parallax (mas) | distance (pc) |
|---|---|---|---|---|---|---|---|---|---|
| | *B* | *V* | *R$_c$* | *I$_c$* | *J* | *H* | *K* | *p* | *d* |
| **TYC 4091-1024-1** | 0.876 | 0.663 | 0.524 | 0.364 | 0.171 | 0.109 | 0.073 | 3.3598 ± 0.0258 | 298 ± 2 |
| **GSC 04091-01022** | 0.825 | 0.624 | 0.494 | 0.342 | 0.161 | 0.102 | 0.069 | 0.9618 ± 0.0160 | 1040 ± 18 |
| **GSC 04091-01031** | 0.819 | 0.619 | 0.490 | 0.340 | 0.160 | 0.101 | 0.068 | 1.2046 ± 0.0189 | 830 ± 13 |
| **GSC 04091-01076** | 0.795 | 0.602 | 0.476 | 0.330 | 0.156 | 0.098 | 0.066 | 1.6297 ± 0.0288 | 614 ± 11 |
| **TYC 4091-1087-1** | 0.799 | 0.605 | 0.478 | 0.332 | 0.156 | 0.099 | 0.067 | 1.3092 ± 0.0260 | 764 ± 15 |
| **TYC 4091-1171-1** | 0.868 | 0.657 | 0.519 | 0.360 | 0.170 | 0.108 | 0.072 | 8.0844 ± 0.0494 | 124 ± 1 |
| **TYC 4091-1838-1** | 0.900 | 0.681 | 0.539 | 0.374 | 0.176 | 0.111 | 0.075 | 2.0266 ± 0.0281 | 493 ± 7 |
| **GSC 04091-00302** | 0.902 | 0.682 | 0.540 | 0.374 | 0.176 | 0.112 | 0.075 | 3.9247 ± 0.0156 | 255 ± 1 |
| **GSC 04091-01109** | 0.911 | 0.689 | 0.545 | 0.378 | 0.178 | 0.113 | 0.076 | 0.8023 ± 0.0191 | 1246 ± 30 |
| **GSC 04091-01163** | 0.874 | 0.661 | 0.523 | 0.363 | 0.171 | 0.108 | 0.073 | 1.2763 ± 0.0190 | 784 ± 12 |

**Table 15: extinction coefficients used and distance of comparison and additional stars**

| | Intrinsic magnitudes (±3σ) | | | | Intrinsic color indexes (±3σ) | | |
|---|---|---|---|---|---|---|---|
| | *B* | *V* | *$R_c$* | *$I_c$* | *B-V* | *$V-R_c$* | *$V-I_c$* |
| **TYC 4091-1024-1** | 11.93 ± 0.08 | 10.96 ± 0.07 | 10.51 ± 0.10 | 10.11 ± 0.07 | 0.97 ± 0.10 | 0.44 ± 0.12 | 0.84 ± 0.10 |
| **GSC 04091-01022** | 12.97 ± 0.08 | 12.36 ± 0.07 | 12.10 ± 0.10 | 11.82 ± 0.07 | 0.61 ± 0.11 | 0.26 ± 0.12 | 0.54 ± 0.10 |
| **GSC 04091-01031** | 13.93 ± 0.08 | 13.34 ± 0.07 | 13.07 ± 0.10 | 12.80 ± 0.07 | 0.59 ± 0.11 | 0.26 ± 0.12 | 0.54 ± 0.10 |
| **GSC 04091-01076** | 13.28 ± 0.08 | 12.45 ± 0.07 | 12.06 ± 0.10 | 11.62 ± 0.07 | 0.83 ± 0.11 | 0.39 ± 0.12 | 0.83 ± 0.10 |
| **TYC 4091-1087-1** | 12.30 ± 0.08 | 11.74 ± 0.07 | 11.49 ± 0.10 | 11.23 ± 0.07 | 0.56 ± 0.11 | 0.24 ± 0.12 | 0.50 ± 0.10 |
| **TYC 4091-1171-1** | 11.21 ± 0.08 | 10.29 ± 0.07 | 9.84 ± 0.10 | 9.45 ± 0.07 | 0.92 ± 0.11 | 0.45 ± 0.12 | 0.84 ± 0.10 |
| **TYC 4091-1838-1** | 11.93 ± 0.04 | 11.19 ± 0.06 | 10.86 ± 0.06 | 10.57 ± 0.03 | 0.73 ± 0.07 | 0.33 ± 0.08 | 0.63 ± 0.07 |
| **GSC 04091-00302** | 14.58 ± 0.13 | 13.55 ± 0.02 | 12.97 ± 0.03 | 12.46 ± 0.01 | 1.03 ± 0.13 | 0.57 ± 0.04 | 1.08 ± 0.02 |
| **GSC 04091-01109** | 13.51 ± 0.12 | 12.97 ± 0.03 | 12.72 ± 0.03 | 12.49 ± 0.02 | 0.54 ± 0.12 | 0.25 ± 0.04 | 0.48 ± 0.04 |
| **GSC 04091-01163** | 13.97 ± 0.23 | 13.33 ± 0.02 | 13.03 ± 0.04 | 12.74 ± 0.02 | 0.63 ± 0.23 | 0.30 ± 0.04 | 0.60 ± 0.03 |
| | *J* | *H* | *K* | *V-J* | *V-H* | *V-K* | *$M_V$* |
| **TYC 4091-1024-1** | 9.522 ± 0.023 | 9.161 ± 0.026 | 9.055 ± 0.020 | 1.43 ± 0.07 | 1.80 ± 0.07 | 1.90 ± 0.07 | 3.59 ± 0.08 |
| **GSC 04091-01022** | 11.716 ± 0.042 | 11.521 ± 0.048 | 11.373 ± 0.033 | 0.65 ± 0.08 | 0.84 ± 0.08 | 0.99 ± 0.08 | 2.28 ± 0.08 |
| **GSC 04091-01031** | 12.505 ± 0.027 | 12.259 ± 0.032 | 12.249 ± 0.28 | 0.83 ± 0.08 | 1.08 ± 0.08 | 1.09 ± 0.08 | 3.74 ± 0.08 |
| **GSC 04091-01076** | 11.141 ± 0.027 | 10.893 ± 0.030 | 10.792 ± 0.021 | 1.41 ± 0.08 | 1.62 ± 0.08 | 1.70 ± 0.08 | 3.51 ± 0.08 |
| **TYC 4091-1087-1** | 10.909 ± 0.026 | 10.762 ± 0.031 | 10.691 ± 0.019 | 0.83 ± 0.07 | 0.98 ± 0.08 | 1.05 ± 0.07 | 2.32 ± 0.08 |
| **TYC 4091-1171-1** | 8.896 ± 0.020 | 8.497 ± 0.028 | 8.410 ± 0.023 | 1.39 ± 0.07 | 1.79 ± 0.08 | 1.88 ± 0.07 | 4.83 ± 0.08 |
| **TYC 4091-1838-1** | 10.294 ± 0.027 | 10.096 ± 0.031 | 10.044 ± 0.026 | 0.90 ± 0.07 | 1.10 ± 0.07 | 1.15 ± 0.07 | 2.73 ± 0.08 |
| **GSC 04091-00302** | 11.755 ± 0.024 | 11.287 ± 0.028 | 11.197 ± 0.022 | 1.79 ± 0.03 | 2.26 ± 0.03 | 2.35 ± 0.03 | 6.52 ± 0.08 |
| **GSC 04091-01109** | 12.406 ± 0.024 | 12.098 ± 0.029 | 12.007 ± 0.024 | 0.79 ± 0.04 | 1.01 ± 0.04 | 1.06 ± 0.04 | 2.49 ± 0.08 |
| **GSC 04091-01163** | 12.357 ± 0.025 | 12.107 ± 0.029 | 12.089 ± 0.024 | 0.98 ± 0.03 | 1.22 ± 0.04 | 1.24 ± 0.03 | 3.86 ± 0.08 |

**Table 16: intrinsic magnitudes, color indexes and absolute magnitude of comparison and additional stars**

| Estimation based on OATo photometry and JHK magnitudes (Cutri et al. 2003) | |
|---|---|
| TYC 4091-1024-1 | G7 IV ÷ K0 IV |
| GSC 04091-01022 | F3 IV ÷ F6 IV |
| GSC 04091-01031 | wF5 V ÷ F6 V |
| GSC 04091-01076 | G5 IV ÷ G8 IV |
| TYC 4091-1087-1 | F5 IV |
| TYC 4091-1171-1 | K0 V |
| TYC 4091-1838-1 | F7 IV ÷ G0 IV |
| GSC 04091-00302 | K3 V |
| GSC 04091-01109 | F5 IV |
| GSC 04091-01163 | F8 V |

**Table 17: spectral types and class of luminosities of comparison and additional stars**

# 5  Conclusions

Our observations and analysis of available photometric data show that variability of VV Cam and VW Cam cannot be excluded, and further observations are needed.

Both stars seem to show an irregular variability with an amplitude of $0.1^m$ in the filter V. The photometric characteristics of the two stars are consistent with a G5 IV- G6 IV subgiant star for VV Cam and with a F5 III - F8 III giant star for VW Cam.

Based on our photometric observations, spectral types for other stars in the field of VV Cam and VW Cam have been estimated for the first time.

## Acknowledgements
- We acknowledge the use of the 1.52m Cassini Telescope run by INAF-Osservatorio Astronomico di Bologna at Loiano site.
- This activity has made use of the SIMBAD database, operated at CDS, Strasbourg, France.
- This work has made use of data from the European Space Agency (ESA) mission Gaia (https://www.cosmos.esa.int/gaia), processed by the Gaia Data Processing and Analysis Consortium (DPAC, https://www.cosmos.esa.int/web/gaia/dpac/consortium). Funding for the DPAC has been provided by national institutions, in particular the institutions participating in the Gaia Multilateral Agreement.
- This work was carried out in the context of educational and training activities provided by Italian law 'Alternanza Scuola Lavoro', July 13th, 2015 n.107, Art.1, paragraphs 33-43.

# Appendix 1 - BVR$_c$I$_c$ photometric data of VV Cam

| VV Cam ||||||||||
|---|---|---|---|---|---|---|---|---|---|---|
| HJD (2400000 +) | B | Err. (±3σ) | HJD (2400000 +) | V | Err. (±3σ) | HJD (2400000 +) | R$_c$ | Err. (±3σ) | HJD (2400000 +) | I$_c$ | Err. (±3σ) |
| 57730.66441 | 13.65 | 0.09 | 57730.66502 | 12.70 | 0.08 | 57730.66551 | 12.24 | 0.10 | 57730.66603 | 11.78 | 0.07 |
| 57730.66662 | 13.67 | 0.09 | 57730.66723 | 12.71 | 0.08 | 57730.66772 | 12.24 | 0.10 | 57730.66824 | 11.79 | 0.07 |
| 57730.66886 | 13.67 | 0.09 | 57730.66945 | 12.70 | 0.08 | 57730.66995 | 12.25 | 0.10 | 57730.67047 | 11.78 | 0.07 |
| 57730.67108 | 13.65 | 0.09 | 57730.67169 | 12.70 | 0.08 | 57730.67218 | 12.24 | 0.10 | 57730.67270 | 11.79 | 0.07 |
| 57730.67330 | 13.66 | 0.09 | 57730.67392 | 12.71 | 0.08 | 57730.67443 | 12.24 | 0.10 | 57730.67495 | 11.78 | 0.07 |
| 57730.67558 | 13.66 | 0.09 | 57730.67620 | 12.70 | 0.08 | 57730.67669 | 12.24 | 0.10 | 57730.67724 | 11.78 | 0.07 |
| 57730.67785 | 13.66 | 0.09 | 57730.67846 | 12.71 | 0.08 | 57730.67896 | 12.24 | 0.10 | 57730.67949 | 11.77 | 0.07 |
| 57730.68011 | 13.66 | 0.09 | 57730.68073 | 12.71 | 0.08 | 57730.68123 | 12.23 | 0.10 | 57730.68176 | 11.78 | 0.07 |
| 57730.68238 | 13.66 | 0.09 | 57730.68300 | 12.70 | 0.08 | 57730.68351 | 12.24 | 0.10 | 57730.68404 | 11.78 | 0.07 |
| 57730.68466 | 13.67 | 0.09 | 57730.68529 | 12.70 | 0.08 | 57730.68579 | 12.24 | 0.10 | 57730.68632 | 11.78 | 0.07 |
| 57730.68694 | 13.66 | 0.09 | 57730.68757 | 12.70 | 0.08 | 57730.68807 | 12.25 | 0.10 | 57730.68862 | 11.78 | 0.07 |
| 57730.68924 | 13.65 | 0.09 | 57730.68986 | 12.70 | 0.08 | 57730.69037 | 12.24 | 0.10 | 57730.69091 | 11.78 | 0.07 |
| 57730.69152 | 13.65 | 0.09 | 57730.69215 | 12.71 | 0.08 | 57730.69267 | 12.24 | 0.10 | 57730.69321 | 11.78 | 0.07 |
| 57730.69384 | 13.65 | 0.09 | 57730.69447 | 12.71 | 0.08 | 57730.69497 | 12.25 | 0.10 | 57730.69553 | 11.78 | 0.07 |
| 57730.69614 | 13.66 | 0.09 | 57730.69677 | 12.70 | 0.08 | 57730.69728 | 12.25 | 0.10 | 57730.69783 | 11.77 | 0.07 |
| 57730.69845 | 13.65 | 0.09 | 57730.69908 | 12.70 | 0.08 | 57730.69961 | 12.24 | 0.10 | 57730.70015 | 11.78 | 0.07 |
| 57730.70078 | 13.66 | 0.09 | 57730.70142 | 12.71 | 0.08 | 57730.70194 | 12.24 | 0.10 | 57730.70248 | 11.78 | 0.07 |
| 57730.70312 | 13.66 | 0.09 | 57730.70375 | 12.70 | 0.08 | 57730.70426 | 12.24 | 0.10 | 57730.70481 | 11.78 | 0.07 |
| 57730.70544 | 13.67 | 0.09 | 57730.70609 | 12.70 | 0.08 | 57730.70661 | 12.23 | 0.10 | 57730.70716 | 11.78 | 0.07 |
| 57730.70781 | 13.66 | 0.09 | 57730.70844 | 12.71 | 0.08 | 57730.70897 | 12.24 | 0.10 | 57730.70954 | 11.78 | 0.07 |
| 57730.71018 | 13.66 | 0.09 | 57730.71082 | 12.70 | 0.08 | 57730.71135 | 12.24 | 0.10 | 57730.71190 | 11.78 | 0.07 |
| 57730.71255 | 13.65 | 0.09 | 57730.71319 | 12.70 | 0.08 | 57730.71373 | 12.24 | 0.10 | 57730.71429 | 11.78 | 0.07 |
| 57730.71493 | 13.66 | 0.09 | 57730.71558 | 12.70 | 0.08 | 57730.71612 | 12.23 | 0.10 | 57730.71666 | 11.78 | 0.07 |
| 57730.71731 | 13.65 | 0.09 | 57730.71794 | 12.70 | 0.08 | 57730.71848 | 12.23 | 0.10 | 57730.71903 | 11.78 | 0.07 |
| 57730.71967 | 13.66 | 0.09 | 57730.72030 | 12.70 | 0.08 | 57730.72083 | 12.24 | 0.10 | 57730.72138 | 11.78 | 0.07 |
| 57730.72202 | 13.64 | 0.09 | 57730.72265 | 12.70 | 0.08 | 57730.72318 | 12.24 | 0.10 | 57730.72372 | 11.78 | 0.07 |
| 57730.72436 | 13.68 | 0.09 | 57730.72500 | 12.69 | 0.08 | 57730.72553 | 12.24 | 0.10 | 57730.72607 | 11.79 | 0.07 |
| 57730.72672 | 13.66 | 0.09 | 57730.72736 | 12.70 | 0.08 | 57730.72789 | 12.24 | 0.10 | 57730.72844 | 11.79 | 0.07 |
| 57730.72908 | 13.67 | 0.09 | 57730.72971 | 12.71 | 0.08 | 57730.73024 | 12.24 | 0.10 | 57730.73078 | 11.78 | 0.07 |
| 57730.73142 | 13.66 | 0.09 | 57730.73206 | 12.70 | 0.08 | 57730.73259 | 12.25 | 0.10 | 57730.73314 | 11.77 | 0.07 |
| HJD (2457730 +) | B-V | Err. (±3σ) | HJD (2457730 +) | V-R$_c$ | Err. (±3σ) | HJD (2457730 +) | V-I$_c$ | Err. (±3σ) | | | |
| 0.66472 | 0.95 | 0.11 | 0.66527 | 0.46 | 0.12 | 0.66553 | 0.92 | 0.10 | | | |
| 0.66693 | 0.96 | 0.11 | 0.66748 | 0.46 | 0.12 | 0.66774 | 0.92 | 0.10 | | | |
| 0.66916 | 0.96 | 0.11 | 0.66970 | 0.46 | 0.12 | 0.66996 | 0.92 | 0.10 | | | |
| 0.67138 | 0.95 | 0.11 | 0.67193 | 0.46 | 0.12 | 0.67219 | 0.92 | 0.10 | | | |
| 0.67361 | 0.95 | 0.11 | 0.67417 | 0.47 | 0.12 | 0.67443 | 0.93 | 0.10 | | | |
| 0.67589 | 0.96 | 0.11 | 0.67645 | 0.46 | 0.12 | 0.67672 | 0.92 | 0.10 | | | |
| 0.67816 | 0.95 | 0.11 | 0.67871 | 0.47 | 0.12 | 0.67898 | 0.93 | 0.10 | | | |
| 0.68042 | 0.95 | 0.11 | 0.68098 | 0.47 | 0.12 | 0.68125 | 0.92 | 0.10 | | | |
| 0.68269 | 0.97 | 0.11 | 0.68325 | 0.46 | 0.12 | 0.68352 | 0.92 | 0.10 | | | |
| 0.68497 | 0.97 | 0.11 | 0.68554 | 0.46 | 0.12 | 0.68581 | 0.92 | 0.10 | | | |
| 0.68725 | 0.96 | 0.11 | 0.68782 | 0.45 | 0.12 | 0.68809 | 0.92 | 0.10 | | | |
| 0.68955 | 0.95 | 0.11 | 0.69012 | 0.46 | 0.12 | 0.69038 | 0.92 | 0.10 | | | |
| 0.69184 | 0.94 | 0.11 | 0.69241 | 0.46 | 0.12 | 0.69268 | 0.93 | 0.10 | | | |
| 0.69415 | 0.94 | 0.11 | 0.69472 | 0.46 | 0.12 | 0.69500 | 0.93 | 0.10 | | | |
| 0.69645 | 0.96 | 0.11 | 0.69703 | 0.46 | 0.12 | 0.69730 | 0.93 | 0.10 | | | |
| 0.69877 | 0.95 | 0.11 | 0.69935 | 0.46 | 0.12 | 0.69962 | 0.92 | 0.10 | | | |
| 0.70110 | 0.95 | 0.11 | 0.70168 | 0.46 | 0.12 | 0.70195 | 0.93 | 0.10 | | | |
| 0.70343 | 0.96 | 0.11 | 0.70401 | 0.46 | 0.12 | 0.70428 | 0.92 | 0.10 | | | |
| 0.70577 | 0.96 | 0.11 | 0.70635 | 0.47 | 0.12 | 0.70662 | 0.92 | 0.10 | | | |
| 0.70812 | 0.95 | 0.11 | 0.70871 | 0.47 | 0.12 | 0.70899 | 0.92 | 0.10 | | | |
| 0.71050 | 0.95 | 0.11 | 0.71108 | 0.46 | 0.12 | 0.71136 | 0.92 | 0.10 | | | |
| 0.71287 | 0.95 | 0.11 | 0.71346 | 0.46 | 0.12 | 0.71374 | 0.91 | 0.10 | | | |
| 0.71526 | 0.96 | 0.11 | 0.71585 | 0.47 | 0.12 | 0.71612 | 0.92 | 0.10 | | | |
| 0.71762 | 0.95 | 0.11 | 0.71821 | 0.47 | 0.12 | 0.71849 | 0.92 | 0.10 | | | |
| 0.71998 | 0.96 | 0.11 | 0.72056 | 0.46 | 0.12 | 0.72084 | 0.92 | 0.10 | | | |
| 0.72233 | 0.95 | 0.11 | 0.72291 | 0.45 | 0.12 | 0.72318 | 0.92 | 0.10 | | | |
| 0.72468 | 0.98 | 0.11 | 0.72526 | 0.45 | 0.12 | 0.72553 | 0.91 | 0.10 | | | |
| 0.72704 | 0.96 | 0.11 | 0.72762 | 0.46 | 0.12 | 0.72790 | 0.91 | 0.10 | | | |
| 0.72939 | 0.96 | 0.11 | 0.72997 | 0.48 | 0.12 | 0.73024 | 0.93 | 0.10 | | | |
| 0.73174 | 0.96 | 0.11 | 0.73232 | 0.45 | 0.12 | 0.73260 | 0.93 | 0.10 | | | |

# Appendix 2 - BVR$_c$I$_c$ photometric data of VW Cam

| VW Cam |||||||||||
|---|---|---|---|---|---|---|---|---|---|---|
| HJD (2400000 +) | B | Err. (±3σ) | HJD (2400000 +) | V | Err. (±3σ) | HJD (2400000 +) | R$_c$ | Err. (±3σ) | HJD (2400000 +) | I$_c$ | Err. (±3σ) |
| 57730.66441 | 14.10 | 0.09 | 57730.66502 | 12.90 | 0.08 | 57730.66551 | 12.32 | 0.10 | 57730.66603 | 11.73 | 0.08 |
| 57730.66662 | 14.10 | 0.09 | 57730.66723 | 12.90 | 0.08 | 57730.66772 | 12.32 | 0.10 | 57730.66824 | 11.73 | 0.08 |
| 57730.66886 | 14.10 | 0.09 | 57730.66945 | 12.89 | 0.08 | 57730.66995 | 12.31 | 0.10 | 57730.67047 | 11.72 | 0.08 |
| 57730.67108 | 14.09 | 0.09 | 57730.67169 | 12.90 | 0.08 | 57730.67218 | 12.32 | 0.10 | 57730.67270 | 11.73 | 0.08 |
| 57730.67330 | 14.10 | 0.09 | 57730.67392 | 12.90 | 0.08 | 57730.67443 | 12.32 | 0.10 | 57730.67495 | 11.73 | 0.08 |
| 57730.67558 | 14.09 | 0.09 | 57730.67620 | 12.89 | 0.08 | 57730.67669 | 12.31 | 0.10 | 57730.67724 | 11.73 | 0.08 |
| 57730.67785 | 14.08 | 0.09 | 57730.67846 | 12.91 | 0.08 | 57730.67896 | 12.31 | 0.10 | 57730.67949 | 11.72 | 0.08 |
| 57730.68011 | 14.09 | 0.09 | 57730.68073 | 12.90 | 0.08 | 57730.68123 | 12.31 | 0.10 | 57730.68176 | 11.73 | 0.08 |
| 57730.68238 | 14.06 | 0.09 | 57730.68300 | 12.90 | 0.08 | 57730.68351 | 12.32 | 0.10 | 57730.68404 | 11.73 | 0.08 |
| 57730.68466 | 14.10 | 0.09 | 57730.68529 | 12.89 | 0.08 | 57730.68579 | 12.32 | 0.10 | 57730.68632 | 11.73 | 0.08 |
| 57730.68694 | 14.10 | 0.09 | 57730.68757 | 12.89 | 0.08 | 57730.68807 | 12.32 | 0.10 | 57730.68862 | 11.73 | 0.08 |
| 57730.68924 | 14.10 | 0.09 | 57730.68986 | 12.90 | 0.08 | 57730.69037 | 12.32 | 0.10 | 57730.69091 | 11.73 | 0.08 |
| 57730.69152 | 14.09 | 0.09 | 57730.69215 | 12.90 | 0.08 | 57730.69267 | 12.31 | 0.10 | 57730.69321 | 11.74 | 0.08 |
| 57730.69384 | 14.09 | 0.09 | 57730.69447 | 12.90 | 0.08 | 57730.69497 | 12.31 | 0.10 | 57730.69553 | 11.72 | 0.08 |
| 57730.69614 | 14.09 | 0.09 | 57730.69677 | 12.90 | 0.08 | 57730.69728 | 12.32 | 0.10 | 57730.69783 | 11.72 | 0.08 |
| 57730.69845 | 14.10 | 0.09 | 57730.69908 | 12.89 | 0.08 | 57730.69961 | 12.31 | 0.10 | 57730.70015 | 11.73 | 0.08 |
| 57730.70078 | 14.09 | 0.09 | 57730.70142 | 12.90 | 0.08 | 57730.70194 | 12.32 | 0.10 | 57730.70248 | 11.73 | 0.08 |
| 57730.70312 | 14.09 | 0.09 | 57730.70375 | 12.90 | 0.08 | 57730.70426 | 12.33 | 0.10 | 57730.70481 | 11.72 | 0.08 |
| 57730.70544 | 14.09 | 0.09 | 57730.70609 | 12.90 | 0.08 | 57730.70661 | 12.31 | 0.10 | 57730.70716 | 11.73 | 0.08 |
| 57730.70781 | 14.09 | 0.09 | 57730.70844 | 12.89 | 0.08 | 57730.70897 | 12.31 | 0.10 | 57730.70954 | 11.73 | 0.08 |
| 57730.71018 | 14.10 | 0.09 | 57730.71082 | 12.90 | 0.08 | 57730.71135 | 12.31 | 0.10 | 57730.71190 | 11.73 | 0.08 |
| 57730.71255 | 14.11 | 0.09 | 57730.71319 | 12.89 | 0.08 | 57730.71373 | 12.31 | 0.10 | 57730.71429 | 11.73 | 0.08 |
| 57730.71493 | 14.11 | 0.09 | 57730.71558 | 12.88 | 0.08 | 57730.71612 | 12.33 | 0.10 | 57730.71666 | 11.74 | 0.08 |
| 57730.71731 | 14.10 | 0.09 | 57730.71794 | 12.89 | 0.08 | 57730.71848 | 12.31 | 0.10 | 57730.71903 | 11.73 | 0.08 |
| 57730.71967 | 14.09 | 0.09 | 57730.72030 | 12.91 | 0.08 | 57730.72083 | 12.31 | 0.10 | 57730.72138 | 11.72 | 0.08 |
| 57730.72202 | 14.09 | 0.09 | 57730.72265 | 12.89 | 0.08 | 57730.72318 | 12.31 | 0.10 | 57730.72372 | 11.72 | 0.08 |
| 57730.72436 | 14.08 | 0.09 | 57730.72500 | 12.91 | 0.08 | 57730.72553 | 12.31 | 0.10 | 57730.72607 | 11.73 | 0.08 |
| 57730.72672 | 14.08 | 0.09 | 57730.72736 | 12.89 | 0.08 | 57730.72789 | 12.32 | 0.10 | 57730.72844 | 11.73 | 0.08 |
| 57730.72908 | 14.09 | 0.09 | 57730.72971 | 12.89 | 0.08 | 57730.73024 | 12.32 | 0.10 | 57730.73078 | 11.74 | 0.08 |
| 57730.73142 | 14.12 | 0.09 | 57730.73206 | 12.90 | 0.08 | 57730.73259 | 12.32 | 0.10 | 57730.73314 | 11.73 | 0.08 |
| HJD (2457730 +) | B-V | Err. (±3σ) | HJD (2457730 +) | V-R$_c$ | Err. (±3σ) | HJD (2457730 +) | V-I$_c$ | Err. (±3σ) | | | |
| 0.66472 | 1.19 | 0.11 | 0.66527 | 0.58 | 0.13 | 0.66553 | 1.16 | 0.11 | | | |
| 0.66693 | 1.20 | 0.11 | 0.66748 | 0.58 | 0.13 | 0.66774 | 1.17 | 0.11 | | | |
| 0.66916 | 1.21 | 0.11 | 0.66970 | 0.58 | 0.13 | 0.66996 | 1.17 | 0.11 | | | |
| 0.67138 | 1.19 | 0.11 | 0.67193 | 0.59 | 0.13 | 0.67219 | 1.17 | 0.11 | | | |
| 0.67361 | 1.20 | 0.11 | 0.67417 | 0.58 | 0.13 | 0.67443 | 1.17 | 0.11 | | | |
| 0.67589 | 1.19 | 0.11 | 0.67645 | 0.59 | 0.13 | 0.67672 | 1.16 | 0.11 | | | |
| 0.67816 | 1.17 | 0.11 | 0.67871 | 0.60 | 0.13 | 0.67898 | 1.19 | 0.11 | | | |
| 0.68042 | 1.19 | 0.11 | 0.68098 | 0.58 | 0.13 | 0.68125 | 1.16 | 0.11 | | | |
| 0.68269 | 1.16 | 0.11 | 0.68325 | 0.58 | 0.13 | 0.68352 | 1.16 | 0.11 | | | |
| 0.68497 | 1.22 | 0.11 | 0.68554 | 0.57 | 0.13 | 0.68581 | 1.16 | 0.11 | | | |
| 0.68725 | 1.21 | 0.11 | 0.68782 | 0.57 | 0.13 | 0.68809 | 1.16 | 0.11 | | | |
| 0.68955 | 1.21 | 0.11 | 0.69012 | 0.57 | 0.13 | 0.69038 | 1.16 | 0.11 | | | |
| 0.69184 | 1.19 | 0.11 | 0.69241 | 0.58 | 0.13 | 0.69268 | 1.16 | 0.11 | | | |
| 0.69415 | 1.19 | 0.11 | 0.69472 | 0.59 | 0.13 | 0.69500 | 1.18 | 0.11 | | | |
| 0.69645 | 1.19 | 0.11 | 0.69703 | 0.58 | 0.13 | 0.69730 | 1.17 | 0.11 | | | |
| 0.69877 | 1.21 | 0.11 | 0.69935 | 0.58 | 0.13 | 0.69962 | 1.16 | 0.11 | | | |
| 0.70110 | 1.19 | 0.11 | 0.70168 | 0.57 | 0.13 | 0.70195 | 1.17 | 0.11 | | | |
| 0.70343 | 1.19 | 0.11 | 0.70401 | 0.57 | 0.13 | 0.70428 | 1.17 | 0.11 | | | |
| 0.70577 | 1.20 | 0.11 | 0.70635 | 0.58 | 0.13 | 0.70662 | 1.16 | 0.11 | | | |
| 0.70812 | 1.20 | 0.11 | 0.70871 | 0.58 | 0.13 | 0.70899 | 1.16 | 0.11 | | | |
| 0.71050 | 1.20 | 0.11 | 0.71108 | 0.59 | 0.13 | 0.71136 | 1.17 | 0.11 | | | |
| 0.71287 | 1.22 | 0.11 | 0.71346 | 0.58 | 0.13 | 0.71374 | 1.15 | 0.11 | | | |
| 0.71526 | 1.22 | 0.11 | 0.71585 | 0.56 | 0.13 | 0.71612 | 1.15 | 0.11 | | | |
| 0.71762 | 1.21 | 0.11 | 0.71821 | 0.58 | 0.13 | 0.71849 | 1.16 | 0.11 | | | |
| 0.71998 | 1.18 | 0.11 | 0.72056 | 0.59 | 0.13 | 0.72084 | 1.18 | 0.11 | | | |
| 0.72233 | 1.20 | 0.11 | 0.72291 | 0.58 | 0.13 | 0.72318 | 1.18 | 0.11 | | | |
| 0.72468 | 1.18 | 0.11 | 0.72526 | 0.59 | 0.13 | 0.72553 | 1.17 | 0.11 | | | |
| 0.72704 | 1.18 | 0.11 | 0.72762 | 0.58 | 0.13 | 0.72790 | 1.16 | 0.11 | | | |
| 0.72939 | 1.21 | 0.11 | 0.72997 | 0.57 | 0.13 | 0.73024 | 1.15 | 0.11 | | | |
| 0.73174 | 1.22 | 0.11 | 0.73232 | 0.58 | 0.13 | 0.73260 | 1.16 | 0.11 | | | |